\documentclass[12pt]{iopart}
\usepackage{graphicx}
\usepackage{dcolumn}
\usepackage{bm}
\usepackage{color}

\begin{document}

\title[Nondiffracting Accelerating Waves]{Nondiffracting Accelerating Waves: Weber waves and parabolic momentum}

\author{Miguel A. Bandres, B.~M. Rodr\'{\i}guez-Lara}

\address{Instituto Nacional de Astrof\'isica, \'Optica y Electr\'onica\\
Calle Luis Enrique Erro No. 1, Sta. Ma. Tonantzintla, Pue. CP 72840, M\'exico}

\ead{bandres@gmail.com}
\begin{abstract}
Diffraction is one of the universal phenomena of physics, and a way to overcome it has always represented a challenge for physicists. In order to control diffraction, the study of structured waves has become decisive. Here, we present a specific class of nondiffracting spatially accelerating solutions of the Maxwell equations: the Weber waves. These nonparaxial waves propagate along parabolic trajectories while approximately preserving their shape. They are expressed in an analytic closed form and naturally separate in forward and backward propagation. We show that the Weber waves are self-healing, can form periodic breather waves and have a well-defined conserved quantity: the parabolic momentum. We find that our Weber waves for moderate to large values of the parabolic momenta can be described by a modulated Airy function. Because the Weber waves are exact time-harmonic solutions of the wave equation, they have implications for many linear wave systems in nature, ranging from acoustic, electromagnetic and elastic waves to surface waves in fluids and membranes.
\end{abstract}

\maketitle

\section{Introduction}

A paraxial accelerating beam is a shape-preserving light beam whose peak intensity follows a continuous parabolic curve as it propagates in free space. 
The first accelerating beam, the Airy beam, was proposed and observed in 2007 \cite{Airy}. 
By now, it has been proved that the only solutions to the paraxial wave equation with accelerating nondiffractive properties are the one-dimensional Airy beams and the two-dimensional Airy and accelerating parabolic beams \cite{ApA,AcB}. 
 The unique properties of the accelerating beams have led to many intriguing ideas and applications, e.g. particle and cell micromanipulation \cite{Dholakia,Dholakia2}, plasma physics \cite{Plasma}, nonlinear optics \cite{NLO}, plasmonics \cite{Plasmon1,Plasmon2} and micromachining \cite{Dudley2}, among others.

Recently, new nondiffractive accelerating waves called `half a Bessel' waves were theoretically introduced \cite{Segev} and experimentally verified \cite{Dudley}. These waves propagate along a circular trajectory; during a quarter of the circle they are quasi shape-preserving and after this, diffraction broadening takes over and the waves spread out.
The importance of these waves consists in having the same characteristics as the paraxial accelerating beams \cite{Airy,ApA,AcB} but in the nonparaxial regime, i.e. these waves can bend to broader angles. Therefore, the `half a Bessel' waves allow one to extrapolate all the intriguing applications of accelerating beams to the nonparaxial regime, and because these waves are solutions to the wave equation, they have implications for many linear wave systems in nature, ranging from acoustic and elastic waves to surface waves in fluids and membranes. Therefore, a natural question is: are there other accelerating nondiffractive solutions to the wave equation?

In this work, we present new nonparaxial spatially accelerating shape-preserving waves: the Weber waves. These nonparaxial waves propagate along parabolic trajectories while approximately preserving their shape within a range of propagation distances. Our Weber waves, like the `half a Bessel' waves, are self-healing, they can form breather waves, and they are a complete and orthogonal family of waves. The Weber waves naturally separate in forward and backward propagation, and they have an analytic closed-form solution.

The `half a Bessel' waves by construction break the circular symmetry and do not have a well-defined angular momentum. In contrast, we show that the Weber waves have a well-defined conserved quantity: the {\it parabolic momentum}. We found that our Weber waves for moderate to large values of the parabolic momenta can be described by a modulated Airy function.
We demonstrated that the local value of the linear momentum density of the Weber waves flows in parabolic lines and this allows to transfer {\it parabolic momentum} to mechanical systems. Accordingly, schemes and uses equivalent to those found for orbital-angular momentum \cite{Allen} can be proposed for {\it parabolic momentum}.

\section{Weber waves}

To construct new nondiffracting accelerating waves of monochromatic Maxwell's equations in a homogeneous medium as in \cite{Segev}, one just needs to find quasi-nondiffracting accelerating solutions of the two-dimensional Helmholtz equation $\left( \partial _{xx}+\partial _{zz}+k^{2}\right) \psi =0$, where $k$ is the wavenumber. Because the waves of interest are nonsingular and we consider only propagation in free space, there can be no evanescent components. Therefore, the spectrum of all the solutions to this equation lies on a circle or radius $k=\omega/c$ in the $k_x-k_z$ plane and it is characterized by the spectral function 
$A\left( \theta \right) =\{A_{+}\left( \theta \right),A_{-}\left( \theta
\right) \}$. From this spectral function, the wave can be constructed by
\begin{equation}
\psi \left( x,z\right) =\left[\int\limits_{-\pi}^{0}A_{-}\left( \theta \right)+\int\limits_{0}^{\pi}A_{+}\left( \theta \right) \right] e^{ik\left( x\cos \theta
+z\sin \theta \right) }\mathrm{d}\theta.   \label{AngInt}
\end{equation}%

We want forward propagating waves in the $z$-coordinate; therefore we are interested only in solutions with components in the top half $0<\theta<\pi$ of the spectral function, i.e. $A_{-}\left(\theta\right)=0$. In the rest of this work, we will assume forward propagating waves; backward propagating waves can be readily constructed by $A\left(\theta\right)\rightarrow \{0,A_{+}\left(-\theta\right)\}$.
By this argument, the `half a Bessel' waves are defined in \cite{Segev} as half of the spectral function of a Bessel beam, i.e. $A_{+}\left( \theta
\right)=\mathrm{exp}(-im\theta)/\sqrt{\pi}$. The `half a Bessel' waves do not have analytic solution in the $x-z$ space because equation (\ref{AngInt}) for its spectral function cannot be solved analytically.
\begin{figure}[t]
\centerline{\includegraphics[width=300pt]{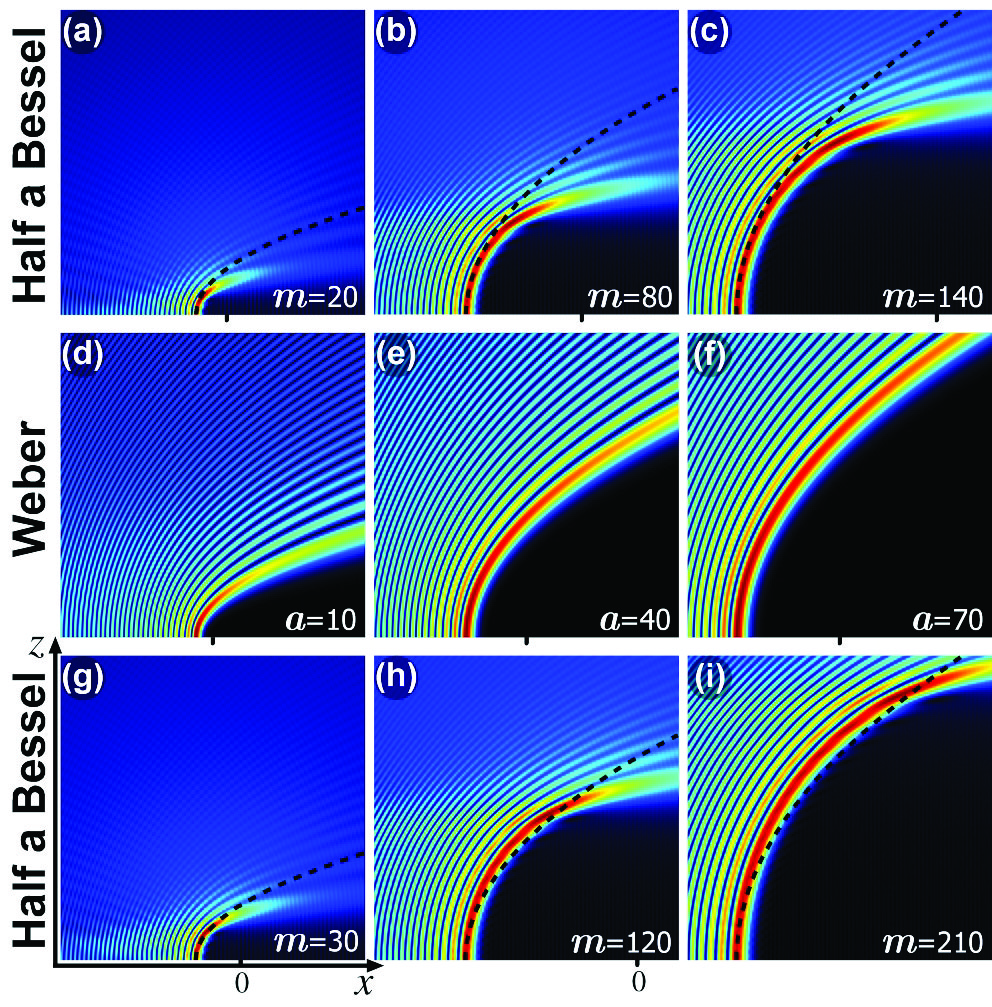}}
\caption{Top row (a-c) and bottom row (g-i): `half a Bessel' waves; middle row (d-f): Weber waves. The dashed black parabolas describe the trajectory of the maximum lobe of the Weber wave in the same column. All figures depict a $220k^{-1}\times220k^{-1}$ square.}\label{Fig1}
\end{figure}

Because we want solutions that have a parabolic translation upon propagation, we look at the two-dimensional Helmholtz equation in parabolic coordinates $x+iz =\left( \eta +i\xi \right) ^{2}/2$ with ranges in $\eta  \in \left( -\infty ,\infty \right)$, $\xi  \in \left[ 0,\infty \right)$. In \cite{Bandres}, a solution whose spectral function just lies on the top half of the spectral circle is created by superposing two parabolic nondiffractive beams. There, this solution was considered as a nondiffractive wave propagating in a straight line, but here we will consider it as an accelerating wave propagating in a parabolic curve; that is, the $y$-axis in \cite{Bandres} becomes the $z$-axis here. As is shown in \cite{Segev} this simple physical reinterpretation of the solution has profound implications. Therefore, we define the Weber waves as \cite{Bandres,Blas}
\begin{equation}
W(\eta ,\xi ;a)=\frac{1}{2 \pi }\left[ 
\begin{array}{c}
\left\vert \Gamma _{1}\right\vert ^{2}\mathrm{P}_{\mathrm{0}}\left( \sigma
\xi ;a\right) \mathrm{P}_{\mathrm{0}}\left( \sigma \eta ;-a\right)  \\ 
+2i\left\vert \Gamma _{3}\right\vert ^{2}\mathrm{P}_{1}\left( \sigma \xi
;a\right) \mathrm{P}_{\mathrm{1}}\left( \sigma \eta ;-a\right) 
\end{array}%
\right] ,\label{WB}
\end{equation}
where $\Gamma_{1}\equiv\Gamma(1/4+ia/2),$
$\Gamma_{3}\equiv\Gamma (3/4+ia/2)$, $\sigma\equiv\sqrt{2k}$, $a$ is a dimensionless parameter of the solution that we will call the {\it parabolic momentum} and $\mathrm{P}_{\mathrm{\gamma}}$ are the parabolic cylinder functions \cite{Abramo} that in terms of the Kummer confluent hypergeometric function $\left. _{1}F_{1}\right.$ can be written as
\begin{equation}
\mathrm{P}_{\mathrm{\gamma }}\left( x,a\right) =x^{\gamma
}e^{-ix^{2}/4}\left. _{1}F_{1}\right. \left[ \frac{1+2\gamma}{4}-\frac{ia}{2}%
;\frac{1+2\gamma}{2};\frac{ix^{2}}{2}\right]. 
\end{equation}
The spectrum of the Weber waves can be obtained from \cite{Bandres,Blas} as
\begin{equation}
A_{+}(\theta;a)  =\frac{1}{\sqrt{2\pi\left\vert \sin\theta\right\vert }%
}\exp\left(  -ia\ln\left\vert \tan\frac{\theta}{2}\right\vert
\right),\label{spectrum}
\end{equation}

For two-dimensional waves the physical inner product, i.e. invariant under Euclidean transformations, is given by $\left\langle \psi _{1}|\psi _{2}\right\rangle =\int_{-\pi }^{\pi}A_{1}^{\ast }\left( \theta \right) A_{2}\left( \theta \right) \mathrm{d}\theta$ \cite{KBW}. With this inner product, the Weber waves are orthogonal with respect to the {\it parabolic momentum}, i.e.,
\begin{equation}
\left\langle W\left( \eta ,\xi ,a\right) |W\left( \eta ,\xi ,a^{\prime
}\right) \right\rangle =\delta \left( a-a^{\prime }\right), 
\end{equation}
where $\delta \left( y \right)$ is the Dirac delta function.
Therefore, the Weber waves form a complete and orthogonal family of forward propagating waves and any forward propagating wave can be represented as a linear superposition of Weber waves with different {\it parabolic momentum} values.

We will now analyze the physical characteristics of our Weber waves. The middle row of figure \ref{Fig1} shows several Weber waves for different values of the parabolic momentum. Note that the Weber waves exhibit a shape-preserving profile while moving on a parabolic trajectory whose shape is controlled by the parabolic momentum. The focal length of the parabola increases with the parabolic momentum. For $a<0$ the Weber waves will propagate to the left because $W(x ,z ;-a)=W(-x ,z ;a)$. The bending angle $\phi_b$ (with respect to the $z$-axis) of a Weber beam is given by $\phi_b=(\pi+\theta)/2$, where $\theta$ is the polar coordinate angle, i.e. $x=r\cos \theta$, $z=r\sin \theta$. We found that the first two lobes of the `half a Bessel' and Weber waves at $z=0$ have the same shape if the parameters of the waves satisfy the relation $m=2a$.
In the top row of figure \ref{Fig1}, we show several `half a Bessel' waves with $m=2a$; although we can see that the `half a Bessel' bends faster, qualitatively, it is possible to see that Weber waves stay true to their propagation invariant nature longer. If we compare the Weber waves with `half a Bessel' waves that remain nondiffractive longer, for example the ones satisfying the relation $m=3a$ depicted in the bottom row of figure \ref{Fig1}, one can see that the bending trajectories are equivalent.

We found, using uniform approximation expansions \cite{Berry,Olver}, that for $a\geq 10$ our Weber waves can be approximated (with an error less than $0.5\%$ relative to the maximum) by
\begin{eqnarray}
W(\eta ,\xi ;a) &\approx \left\vert \Gamma _{1}\right\vert ^{2}\frac{e^{\pi
a/2}}{2\sqrt{\pi }}\left( \frac{1}{\widetilde{\eta }^{2}+1}\right)
^{1/4}e^{i\mathrm{H}\left( \widetilde{\eta }\right) } 
\left( \frac{\Xi \left( \widetilde{%
\xi }\right) }{\widetilde{\xi }^{2}-1}\right) ^{1/4}\mathrm{Ai}\left[
-\Xi \left( \widetilde{\xi }\right) \right], 
\label{Aprox}
\end{eqnarray}%
where $\widetilde{x}=x\sqrt{k/2a}$ and
\begin{equation}
\frac{\Xi\left( x\right)}{\left( 6a\right) ^{2/3}}  =\left\{ 
\begin{array}{c}
-\left( \frac{1}{4}\arccos x-\frac{1}{4}x\sqrt{1-x^{2}}\right) ^{2/3}, \\ 
\left( \frac{1}{4}x\sqrt{x^{2}-1}-\frac{1}{4}\mathrm{arccosh} x\right) ^{2/3},%
\end{array}%
\right. 
\begin{array}{c}
x\leq 1, \\ 
x\geq 1,%
\end{array}
\end{equation}
\begin{equation}
\mathrm{H}\left( x\right) =a\left[x\sqrt{x^{2}+1}+\log \left(
x+\sqrt{x^{2}+1}\right)\right]. 
\end{equation}
This form of the Weber wave unravels its physical characteristics. Note that now our waves separate into a product of functions of the $\eta$ and $\xi$ coordinates. The $\xi$ part describes the profile characteristics, while the $\eta$ part describes the propagation characteristics.  The $\xi$ part is a modulation of the Airy function; the argument of the Airy function makes this part oscillate asymptotically as $\sin\left(k\xi^2/2+\pi/4\right)$, i.e. the maximum frequency of oscillation of the wave will be $k$ as expected because the Weber waves do not contain evanescent waves. The modulation makes the envelope of the function to decay as $\xi^{-1/2}$, i.e. as an Airy beam.

\begin{figure}[tb]
\centerline{\includegraphics[width=300pt]{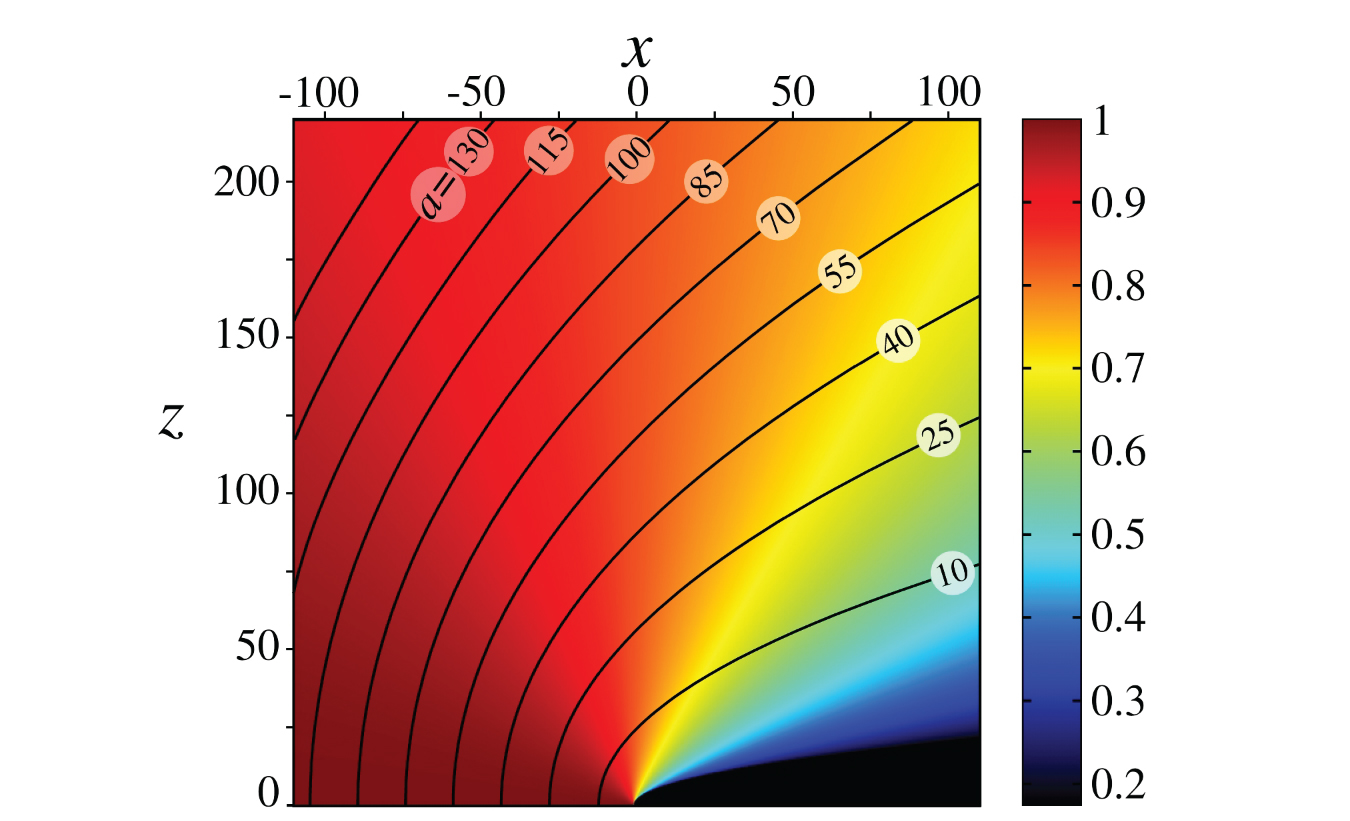}}
\caption{The color map and parabolic lines shows the decay and propagation trajectory, respectively, of the main lobe of the Weber waves for different {\it parabolic momentum} $a$. The coordinates $x$ and $z$ are in units of $k^{-1}.$}
\label{Fig2}
\end{figure}

We found that the maximum lobe of the Weber beam propagates in the parabola with
\begin{equation}
\sqrt{2k}\xi _{\max }\approx 2\sqrt{a}+0.9838a^{-4/25},
\label{xhi_max}
\end{equation}
and by equation (\ref{Aprox}) the value of this maximum will decay upon propagation by $\delta=\left( k\eta^2/2a+1\right) ^{-1/4}$. For Weber waves with $a>10$ this decay can be simplified within an error of $5\%$ by $\delta=\sqrt{\cos\phi_b}$; from this, we see that our waves will bend to $75^\circ$ with still $50\%$ of its peak amplitude.
Figure \ref{Fig2} shows the decay and propagation trajectory  of the main lobe of the Weber waves for different values of {\it parabolic momentum}, i.e. it characterizes the nondiffractive accelerating behavior of our waves.

\begin{figure}[tb]
\centerline{\includegraphics[width=300pt]{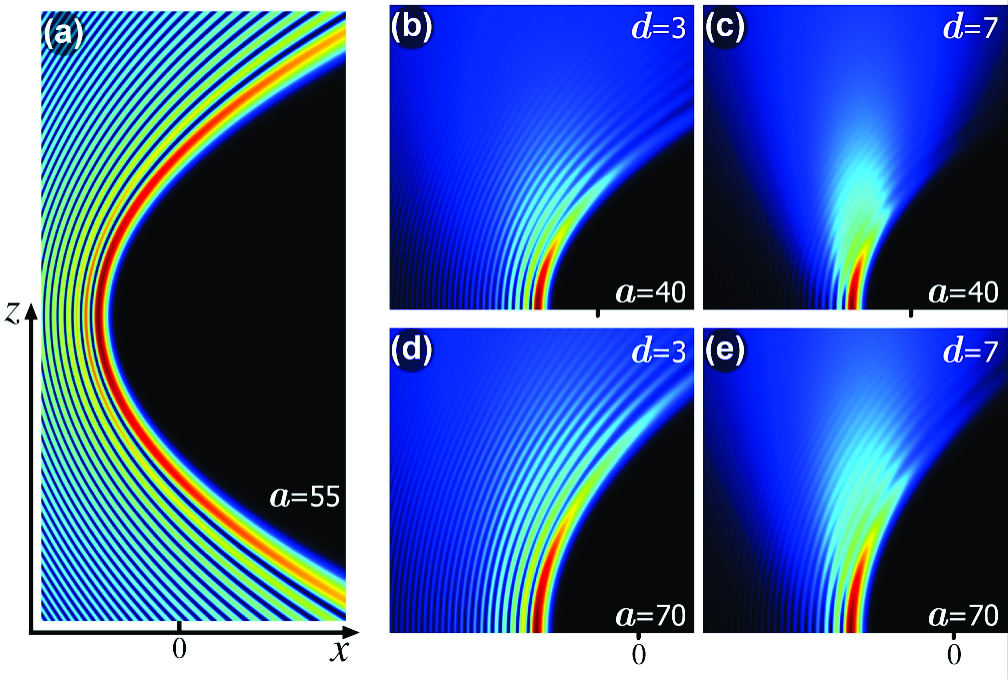}}
\caption{(a) Weber wave $a=55$; (b-e) Paraxial Weber waves with $d=3,7$. (a) depicts a $440k^{-1}\times220k^{-1}$ rectangle and (b-e) all depict a $220k^{-1}\times220k^{-1}$ square.}\label{Fig4}
\end{figure}

It is possible to double the angle of bending by propagating the Weber waves from $z<0$. In this case the wave has a bending angle opposite to the direction of bending and the Weber wave will depict full parabolas as shown in figure \ref{Fig4}(a). These Weber waves can be used for micromachining of micron-size curved structures \cite{Dudley2}.

Because the Weber waves are separable solutions of the wave equation, they are also eigenfunctions of a second-order operator that we will call the {\it parabolic momentum} operator \cite{Miller}. This operator commutes with the Helmholtz equation; therefore, the eigenvalue of the {\it parabolic momentum} operator is a conserved dynamical constant. The {\it parabolic momentum} eigenvalue problem is given by
\begin{equation}
\left( L\partial _{z}+\partial _{z}L\right) W=\frac{(\eta ^{2}\partial _{\xi
\xi }-\xi ^{2}\partial _{\eta \eta })}{\left( \eta ^{2}+\xi ^{2}\right) }%
W=2kaW,
\label{Parabolic_momenta}
\end{equation}%
where $L=x\partial _{z}-z\partial _{x}=\left( \eta \partial _{\xi }-\xi \partial
_{\eta }\right) /2$ is the angular momentum operator.

To explain the physics of the {\it parabolic momentum}, let us calculate the local value of the linear momentum density of the Weber waves $W=\left\vert W\right\vert e^{i\beta}$,
\begin{eqnarray}
\mathbf{p} &=i\omega \varepsilon _{0}\left\vert W\right\vert ^{2}\nabla
\beta, \nonumber  \\
&=i\omega \varepsilon _{0}\left\vert W\right\vert ^{2}k\frac{\sqrt{\eta
^{2}+2a/k}}{\sqrt{\eta ^{2}+\xi ^{2}}}\widehat{\eta}.
\end{eqnarray}

This means that the linear momentum flows purely in parabolas. Therefore, the transverse structure of the Weber waves allows the trapping of micrometer particles by a gradient force, and the linear momentum density will move these particles in parabolic trajectories.  This opens up the possibility of applying schemes and uses equivalent to those found for orbital-angular momentum \cite{Allen} to the {\it parabolic momentum}.

\begin{figure}[t]
\centerline{\includegraphics[width=300pt]{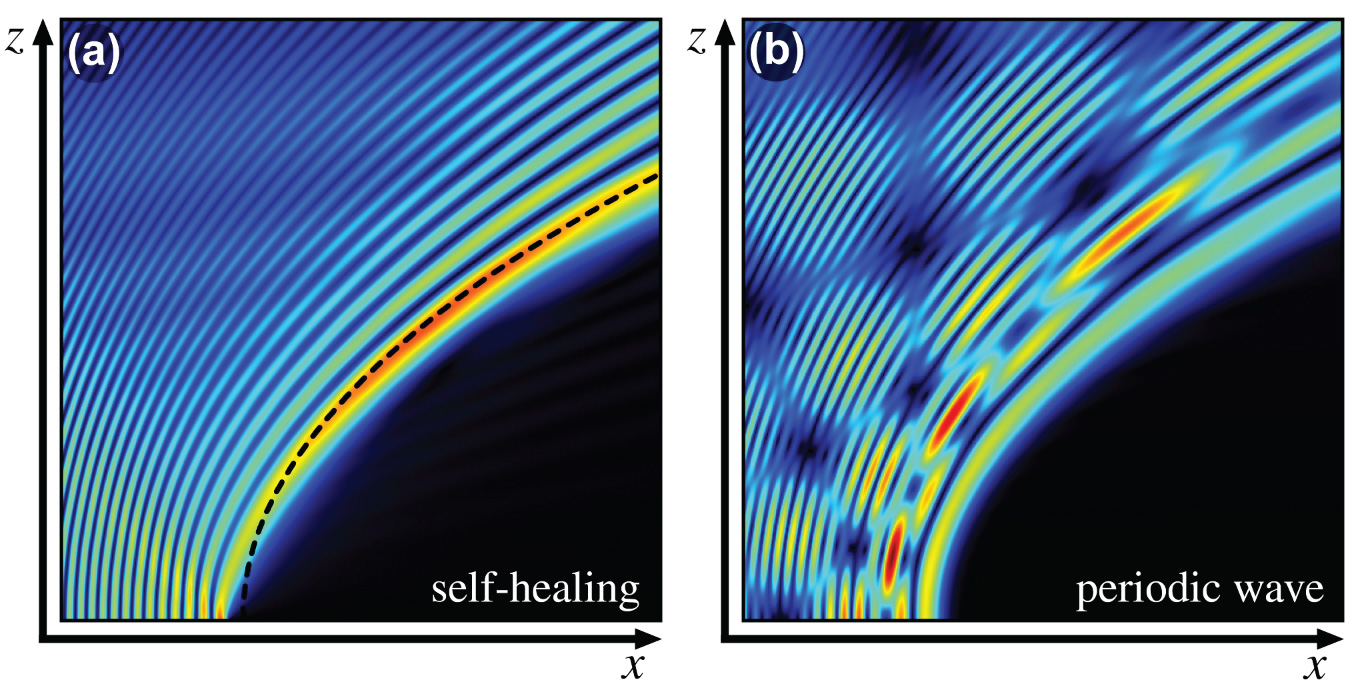}}
\caption{(a) The self-healing property of a Weber wave ($a=40$) with the first lobe initially cut out. The dashed black curved describes the original trajectory of the first lobe. (b) A periodic Weber wave created by supperposing two Weber waves with $a=40$ and $a=54$ with equal coefficientes. All figures depict a $220k^{-1}\times220k^{-1}$ square.} 
\label{Fig3}
\end{figure}

We will now compare the known features of the Airy beam and `half a Bessel' waves with our new Weber waves. First, the Weber waves are also self-healing; for example, in figure \ref{Fig3}(a) the main lobe of a Weber wave is initially blocked and during propagation the second lobe  gets more power and its trajectory bends to replace the first lobe, and similarly each subsequent lobe will bend to replace the lobe on its right.

As in the case of the `half a Bessel' waves, by superposing Weber waves with different values of {\it parabolic momentum}, it is possible to create breather waves that will propagate in a parabolic trajectory but their amplitude will oscillate with periodicity that depends on the difference between the values of the {\it parabolic momentum}, see figure \ref{Fig3}(b).
Similar to the Airy beams and the `half a Bessel' waves, our Weber waves are not square integrable, i.e. they carry infinite power. However, as is common with accelerating beams, if we introduce an aperture or an exponential apodization, the Weber waves will hold their properties but over a smaller distance; the wider the aperture the larger the distance. Also, while the acceleration of the Airy beams can be controlled by an arbitrary constant that scales the $x-$axis linearly and the $z-$axis quadratically, the acceleration of our Weber beams is controlled by $a$, the {\it parabolic momentum}, as explained above.

It is important to note that although Airy beams are solutions to the paraxial equation, the Airy beams do not satisfy the paraxial approximation because their spectrum is not confined to small angles; this is equivalent to the fact that Airy beams have an oscillating tail of constantly increasing frequency. Any solution of the wave equation only satisfies the paraxial equation in the paraxial regime; therefore, it is not possible to recover the Airy beams by varying the parameters of the `half a Bessel' wave or the Weber waves. Furthermore, neither the `half a Bessel' nor the Weber waves, in their current form, has a parameter that controls paraxiality.
In order to control the paraxiality of the waves we can use the idea of imaginary displacement \cite{Sheppard,Alonso}. If one considers a spatial shift by the imaginary distance $\mathrm{i}d/k$ in the $z$-direction, then the spectral function $A(\theta)$ becomes $A(\theta) exp(d\sin\theta)$. Therefore, as $d$ increases, the spectrum becomes increasingly confined around the propagation direction and the field becomes increasingly beam-like. The amplitude of the spectral function of the Weber waves has a singularity at $\theta=0,\pi$; we will call this a soft singularity because as the amplitude diverges, the phase also diverges; therefore, in any experimental or numerical application this divergence will average to zero. Nevertheless, it is possible to cancel this divergence if we subtract the original spectral function from the shifted one, i.e. $A(\theta)\left(exp(d\sin\theta)-1\right)$. In this case, the intensity of the spectral function at $\theta=0,\pi$ relative to the maximum intensity is $d^2/exp(2d)$: consequently, in any practical application we can consider these highly nonparaxial components as zero if $d>3$. Figures \ref{Fig4}(b-e) depicts the Weber waves constructed with this spectral function for $d=3,7$.

The Weber waves can be experimentally generated with the same setup used to generate the `half a Bessel' waves in \cite{Peng2}; in our case the holographic mask has to encode the Fourier transform of equation (\ref{WB}) at $z=0$ which is given by $\left(1-k_{x}^2/k^2\right)^{-1/2} A_+\left(\mathrm{acos}\left(k_{x}/k\right);a\right)$ for $\left\vert k_x\right\vert<k$ and zero otherwise. Transverse electric (TE) and transverse magnetic (TM) vector field solutions of the Maxwell equations can be readily constructed by using our Weber waves as Hertz vector potentials, e.g. defining $\mathbf{\Pi}_{m} = W(\eta,\xi) \hat{z}$ and $\mathbf{\Pi}_{e} = W(\eta,\xi) \hat{z}$. Furthermore, arbitrary polarization vectorial Weber waves are given by equation (5) of \cite{Segev}.

\section{Conclusion}

To summarize, we found nonparaxial accelerating waves that propagate along a parabolic trajectory while preserving their shape to a good approximation. These waves have a well-defined conserved dynamical constant that characterize them, the parabolic momentum. For moderate to large values of the parabolic momenta, the Weber waves can be described by a modulated Airy function; therefore these waves in functional form are the most closely related to the paraxial Airy beams. This work shows that there is another nonparaxial accelerating solution to the wave equation and that the `half a Bessel' waves are not an isolated case. This opens up the possibility of exploring new accelerating solutions of the Maxwell equations. 
Finally, because the Weber waves are the exact time-harmonic solution of the wave equation, they have implications for many linear wave systems in nature, ranging from sound, elastic and surface waves to many kinds of classical waves.


\section*{References}

\begin{thebibliography}{10}

\bibitem{Airy}
G.~A. Siviloglou, J.~Broky, A.~Dogariu, and D.~N. Christodoulides.
\newblock Observation of accelerating airy beams.
\newblock {\em Phys. Rev. Lett.}, 99:213901, Nov 2007.

\bibitem{ApA}
Miguel~A. Bandres.
\newblock Accelerating parabolic beams.
\newblock {\em Optics Letters}, 33(15):1678--1680, August 2008.

\bibitem{AcB}
Miguel~A. Bandres.
\newblock Accelerating beams.
\newblock {\em Opt. Lett.}, 34(24):3791--3793, Dec 2009.

\bibitem{Dholakia}
Jorg Baumgartl, Michael Mazilu, and Kishan Dholakia.
\newblock Optically mediated particle clearing using airy wavepackets.
\newblock {\em Nature Photonics}, 2(11):675--678, September 2008.

\bibitem{Dholakia2}
Jorg Baumgartl, Gregor~M. Hannappel, David~J. Stevenson, Daniel Day, Min Gu,
  and Kishan Dholakia.
\newblock Optical redistribution of microparticles and cells between
  microwells.
\newblock {\em Lab Chip}, 9:1334--1336, 2009.

\bibitem{Plasma}
Pavel Polynkin, Miroslav Kolesik, Jerome~V. Moloney, Georgios~A. Siviloglou,
  and Demetrios~N. Christodoulides.
\newblock Curved plasma channel generation using ultraintense airy beams.
\newblock {\em Science}, 324(5924):229--232, 2009.

\bibitem{NLO}
Tal Ellenbogen, Noa {Voloch-Bloch}, Ayelet {Ganany-Padowicz}, and Ady Arie.
\newblock Nonlinear generation and manipulation of airy beams.
\newblock {\em Nature Photonics}, 3(7):395--398, June 2009.

\bibitem{Plasmon1}
Alexander Minovich, Angela Klein, Norik Janunts, Thomas Pertsch, Dragomir
  Neshev, and Yuri Kivshar.
\newblock Generation and {Near-Field} imaging of airy surface plasmons.
\newblock {\em Physical Review Letters}, 107(11):116802, September 2011.

\bibitem{Plasmon2}
L.~Li, T.~Li, S.~Wang, C.~Zhang, and S.~Zhu.
\newblock Plasmonic airy beam generated by {In-Plane} diffraction.
\newblock {\em Physical Review Letters}, 107(12):126804, September 2011.

\bibitem{Dudley2}
A.~Mathis, F.~Courvoisier, L.~Froehly, L.~Furfaro, M.~Jacquot, P.~A. Lacourt,
  and J.~M. Dudley.
\newblock Micromachining along a curve: Femtosecond laser micromachining of
  curved profiles in diamond and silicon using accelerating beams.
\newblock {\em Applied Physics Letters}, 101(7):071110, 2012.

\bibitem{Segev}
Ido Kaminer, Rivka Bekenstein, Jonathan Nemirovsky, and Mordechai Segev.
\newblock Nondiffracting accelerating wave packets of maxwell's equations.
\newblock {\em Phys. Rev. Lett.}, 108:163901, Apr 2012.

\bibitem{Dudley}
F.~Courvoisier, A.~Mathis, L.~Froehly, R.~Giust, L.~Furfaro, P.~A. Lacourt,
  M.~Jacquot, and J.~M. Dudley.
\newblock Sending femtosecond pulses in circles: highly nonparaxial
  accelerating beams.
\newblock {\em Opt. Lett.}, 37(10):1736--1738, May 2012.

\bibitem{Allen}
L.~Allen, S.M. Barnett, and M.J. Padgett.
\newblock {\em Optical angular momentum}.
\newblock Institute of Physics Pub., Bristol; Philadelphia, 2003.

\bibitem{Bandres}
Miguel~A. Bandres, Julio~C. Guti\'{e}rrez-Vega, and Sabino Ch\'{a}vez-Cerda.
\newblock Parabolic nondiffracting optical wave fields.
\newblock {\em Opt. Lett.}, 29(1):44--46, Jan 2004.

\bibitem{Blas}
B.~M. Rodr\'iguez-Lara and R.~J\'auregui.
\newblock Dynamical constants of structured photons with parabolic-cylindrical
  symmetry.
\newblock {\em Phys. Rev. A}, 79:055806, May 2009.

\bibitem{Abramo}
Milton Abramowitz.
\newblock {\em Handbook of Mathematical Functions}.
\newblock Dover, New York, 1972.

\bibitem{KBW}
Stanly Steinberg and Kurt~Bernardo Wolf.
\newblock Invariant inner products on spaces of solutions of the {Klein–Gordon}
  and helmholtz equations.
\newblock {\em Journal of Mathematical Physics}, 22(8):1660--1663, August 1981.

\bibitem{Berry}
M.~V. Berry.
\newblock Uniform approximation: a new concept in wave theory.
\newblock{\em Sci. Prog.}, 57:43--64, 1969.

\bibitem{Olver} 
F.~W.~J. Olver
\newblock Uniform asymptotic expansions for Weber parabolic cylinder functions of large orders.
\newblock {\em J. Res. Bur. Stand.}, 63B(2):131--173, 1959.

\bibitem{Miller}
C.~P. Boyer, E.~G. Kalnins, and W.~Miller.
\newblock Symmetry and separation of variables for the helmholtz and laplace
  equations.
\newblock {\em Nagoya Mathematical Journal}, 60:35--80, 1976.

\bibitem{Sheppard}
C.~J.~R. Sheppard and S.~Saghafi.
\newblock Beam modes beyond the paraxial approximation: A scalar treatment.
\newblock {\em Phys. Rev. A}, 57:2971--2979, Apr 1998.

\bibitem{Alonso}
Nicole~J. Moore and Miguel~A. Alonso.
\newblock Bases for the description of monochromatic, strongly focused, scalar
  fields.
\newblock {\em J. Opt. Soc. Am. A}, 26(7):1754--1761, Jul 2009.

\bibitem{Peng2}
Peng Zhang, Yi~Hu, Drake Cannan, Alessandro Salandrino, Tongcang Li, Roberto
  Morandotti, Xiang Zhang, and Zhigang Chen.
\newblock Generation of linear and nonlinear nonparaxial accelerating beams.
\newblock {\em Opt. Lett.}, 37(14):2820--2822, Jul 2012.

\end{thebibliography}
\end{document}